\documentclass{article}
\usepackage[preprint]{neurips_2026}

\usepackage[utf8]{inputenc}
\usepackage[T1]{fontenc}
\usepackage{hyperref}
\usepackage{url}
\usepackage{booktabs}
\usepackage{amsfonts}
\usepackage{nicefrac}
\usepackage{microtype}
\usepackage{xcolor}
\usepackage{graphicx}
\usepackage{amsmath}
\usepackage{tikz}

\title{Bipartite Cholesky Graph Networks for Many-Body Quantum Chemistry}

\author{
  Abdul Samad Khan \\
  Lahore University of Management Sciences \\
  \texttt{24120006@lums.edu.pk} \\
}

\begin{document}
\maketitle

\begin{abstract}
Accurate prediction of molecular correlation energies from first principles
requires resolving the $\mathcal{O}(N^4)$ electron repulsion integral (ERI)
tensor. Existing graph neural network approaches to the electronic structure
problem often compress this tensor into low-rank scalar features, discarding
higher-order interaction structures relevant to electron correlation. In this
work, we demonstrate that tensor factorization of the ERI naturally induces a
structured bipartite message-passing architecture that preserves access to
higher-order interaction structure more effectively than compressed orbital
representations. By utilizing the density-fitted Cholesky decomposition of
the ERI tensor, we derive a bipartite graph network that models orbital degrees
of freedom and auxiliary interaction nodes as distinct sets, maintaining
interaction topology at a reduced theoretical complexity of $\mathcal{O}(N^3)$.
Evaluated on 132 geometries of six diatomic molecules with Full Configuration
Interaction (FCI) reference energies, our factorized representation achieves an
in-distribution Mean Absolute Error (MAE) of $0.0296$ Ha under five-fold
cross-validation, a substantial improvement over compressed-integral baselines.
Leave-one-molecule-out validation reveals that zero-shot generalization varies
by nearly a factor of four across molecular species and correlates with the
structural similarity of the held-out molecule's orbital environment to the
training distribution, rather than with nuclear charge asymmetry alone.
\end{abstract}

\section{Introduction}

A central challenge in quantum chemistry is the prediction of molecular
ground-state energies, which requires solving the electronic structure problem
(ESP). While Full Configuration Interaction (FCI) provides exact solutions within
a defined basis set, its exponential scaling renders it computationally
intractable beyond minimal systems. Machine learning surrogates, particularly
Graph Neural Networks (GNNs), bypass explicit ESP solvers by learning mappings
from molecular representations to ground-state properties.

Recent approaches map molecular orbitals to graph nodes, utilizing one- and
two-electron integrals as geometric features. However, to bypass the
$\mathcal{O}(N^4)$ representational bottleneck of the two-electron repulsion
integral (ERI) tensor, existing orbital GNNs frequently compress the ERI tensor
into scalar descriptors such as Coulomb and exchange matrices. This reduction
potentially discards the structural information of pairwise interactions relevant
to electron correlation.

In this work, we demonstrate that tensor factorization of the ERI naturally
induces a structured bipartite message-passing architecture that retains more
higher-order interaction structure than compressed orbital graph representations.
Motivated by the Cholesky factorization of the two-electron tensor,
$g_{pqrs} \approx \sum_L B^L_{pq} B^L_{rs}$, we map the orbital interactions
onto a bipartite contraction graph. This formulation structures the four-index
contraction into a sparse network of orbital nodes exchanging information through
auxiliary interaction channels, and the resulting graph topology is determined
entirely by the mathematical structure of the ERI rather than by heuristic
feature engineering.

Our core contributions are as follows:

\begin{itemize}
    \item We derive a bipartite graph representation motivated by the Cholesky
    factorization of the ERI tensor, bridging tensor-decomposition methods from
    \textit{ab initio} quantum chemistry with orbital-basis deep learning.

    \item We develop a structured message-passing architecture with an empirical
    forward-pass scaling of $\mathcal{O}(N^{2.20})$, well below the
    $\mathcal{O}(N^4)$ cost of explicit ERI evaluation.

    \item We achieve an MAE of $0.0296$ Ha on FCI-computed correlation energy
    targets, improving substantially over compressed-integral baselines under
    identical training conditions.

    \item Through leave-one-molecule-out (LOMO) validation, we show that
    zero-shot generalization error ranges from $0.040$ to $0.161$ Ha across
    species and tracks the orbital-structural similarity of the held-out molecule
    to the training distribution rather than nuclear charge asymmetry alone.
\end{itemize}

\section{Related Work}

\paragraph{Orbital-Basis Machine Learning}
While most chemical machine learning operates in real space, predicting
correlation energies demands a representation grounded in the Hilbert space of
the system. \citet{welborn2018transferability} demonstrated that molecular
orbital (MO) properties---matrix elements of the Fock, Coulomb, and exchange
operators---can be utilized in Gaussian process regression to predict
post-Hartree-Fock correlation energies. This was scaled by
\citet{qiao2020orbnet} in OrbNet, which constructed a GNN using
symmetry-adapted atomic-orbital features. \citet{schutt2019unifying} introduced
SchNOrb, extending deep tensor neural networks to directly predict electronic
wavefunctions in a local orbital basis, enabling the derivation of all
ground-state properties from a single learned representation. More recently,
\citet{chuiko2025predicting} utilized rotationally and unitarily invariant
descriptors built from one- and two-electron integrals, and
\citet{christensen2020fchl} developed the FCHL19 representation offering
chemical accuracy on standard benchmark datasets. However, to avoid the steep
memory footprint of the ERI tensor, these methods generally compress or flatten
the integrals into scalar vectors or heuristic features. While computationally
efficient, this dimension reduction discards the factorized coupling structure
relevant to dynamical electron correlation.

Concurrent to our work, \citet{carrasquilla2025gnn} demonstrated that GNNs
operating on the Fock, Coulomb, and exchange matrices of the molecular orbital
basis can learn FCI energies for the same six-molecule diatomic benchmark,
achieving a per-molecule MAE ranging from $0.066$ to $1.34$ Ha on total energy
targets with an 80/20 geometry split.
Our approach differs in three key respects: (i) the bipartite graph topology is
derived directly from the Cholesky factorization of the ERI tensor rather than
from compressed $J$/$K$ matrices; (ii) we adopt a $\Delta$-ML correlation
energy target that removes the dominant mean-field variance; and (iii) we report
five-fold cross-validation MAE with rigorous out-of-fold prediction tracking
rather than a single held-out split.

\paragraph{Atomistic Message Passing}
The dominant paradigm in geometric deep learning for chemistry maps atoms to
graph nodes and spatial proximity to edges. Foundational frameworks such as
Message Passing Neural Networks (MPNNs) \citep{gilmer2017neural} and SchNet
\citep{schutt2017schnet} rely on continuous-filter convolutions over 3D
coordinates. Recent advancements, such as PAMNet \citep{zhang2023universal},
introduce physics-informed inductive biases to explicitly separate local and
non-local mechanical interactions within macromolecules. While atomistic graphs
excel at modeling classical steric effects and equilibrium geometries, they do
not explicitly encode the non-local electronic interactions formalized in second
quantization, as they observe the distance between nuclei rather than the
interaction vertex mediating the electron density.

\paragraph{Tensor Factorization in Electronic Structure}
Within \textit{ab initio} quantum chemistry, the prohibitive $\mathcal{O}(N^4)$
scaling of the ERI tensor is classically mitigated via density fitting or
Cholesky decomposition. As established by \citet{beebe1977simplifications}, the
positive semi-definiteness of the Coulomb operator allows the ERI supermatrix to
be decomposed into a product of three-index tensors. \citet{koch2003reduced}
demonstrated that this factorization yields substantial computational savings for
correlated methods, and \citet{pedersen2009density} extended the framework to
derive unbiased auxiliary basis sets directly from Cholesky decompositions.
This factorization remains a cornerstone of large-scale post-Hartree-Fock methods
\citep{blaschke2021cholesky}. Our work bridges these domains: rather than
compressing the ERI tensor to fit a standard graph neural network, we use the
Cholesky factorization to structurally define a bipartite geometric graph.

\section{Background}

In second quantization, the non-relativistic electronic Hamiltonian for $N$
spatial orbitals is determined by the one-electron core Hamiltonian $h_{pq}$
and the two-electron repulsion integral (ERI) tensor $g_{pqrs}$. Post-Hartree-Fock
methods calculate the correlation energy by contracting these integrals with
reduced density matrices. The primary computational bottleneck is $g_{pqrs}$,
which dictates an $\mathcal{O}(N^4)$ scaling.

Standard density fitting algorithms mitigate this by projecting the four-index
tensor onto an auxiliary basis of size $N_{aux} \approx 2N$
\citep{koch2003reduced, pedersen2009density}. Mathematically, this corresponds
to an incomplete Cholesky decomposition of the symmetric, positive semi-definite
ERI supermatrix. While prior graph networks have incorporated density-fitted
features, they typically compress the auxiliary dimension to construct static
edge features, losing the multipartite connectivity inherent to the tensor
structure.

\section{Method}

\subsection{Problem Formulation}

We adopt a $\Delta$-machine learning formulation
\citep{ramakrishnan2015big}, explicitly targeting the correlation energy
$\Delta E_{corr} = E_{FCI} - E_{HF}$. The Hartree-Fock mean-field energy
$E_{HF}$ isolates the nuclear repulsion and one-body kinetic contributions,
whose variance across disparate molecular species is $\mathcal{O}(10^2)$
Hartree. The correlation energy $\Delta E_{corr}$, by contrast, spans only
$\mathcal{O}(10^{-1})$ Hartree, and is the physically meaningful quantity
encoding many-body electronic interactions beyond the mean field (see
Appendix~\ref{app:delta_ml} for a detailed justification). We define a
parameterized mapping $f_\theta$ that operates strictly on the one-body
integrals and the factorized interaction vertex:
\[
  f_\theta : (h_{pq},\, B^L_{pq}) \;\to\; \Delta E_{corr}
\]
This isolates the network's objective to the quantum many-body contributions
while removing the dominant mean-field signal from the loss landscape.

\subsection{Cholesky Factorization}

The ERI supermatrix induced by the Coulomb kernel is positive semi-definite.
By mapping orbital pairs $(pq)$ and $(rs)$ to composite indices $I$ and $J$,
we form the symmetric supermatrix $G_{IJ} \equiv g_{pqrs}$. The Cholesky
decomposition approximates this matrix as $G \approx LL^T$. Unpacking the
composite indices yields the factorization:
\[
  g_{pqrs} \;\approx\; \sum_{L=1}^{N_{aux}} B^L_{pq}\, B^L_{rs}
\]
This reduces the parameterization scaling of the interaction space from
$\mathcal{O}(N^4)$ to $\mathcal{O}(N^2 N_{aux})$.

\subsection{Bipartite Interaction Graph}

We translate this factorization into a graph topology
$\mathcal{G} = (\mathcal{V}_O, \mathcal{V}_A, \mathcal{E})$.
The \emph{orbital nodes} $\mathcal{V}_O$ represent the orbital degrees of
freedom ($N$ nodes) with features initialized from the core Hamiltonian:
$\mathbf{x}_p = [h_{pp},\, \|h_p\|_2]$. The \emph{auxiliary interaction nodes}
$\mathcal{V}_A$ ($N_{aux}$ nodes) represent the factorized interaction channels,
initialized to zero. The edges $\mathcal{E}$ connect an orbital pair $(p,q)$
to an auxiliary node $L$ with deterministic weight $B^L_{pq}$. This topology is
fixed by the chemistry and carries no learned parameters, encoding the physical
interaction structure directly as graph connectivity.

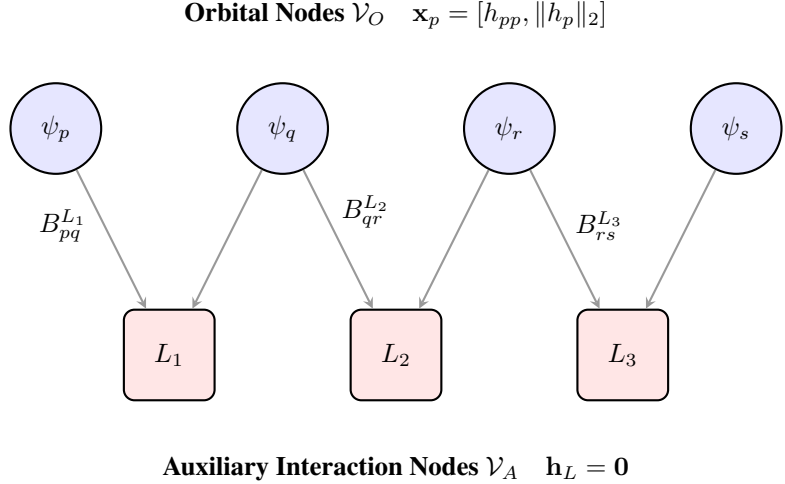
\begin{figure}[ht]
    \centering
    \begin{tikzpicture}[>=stealth]
        \node[circle, draw, thick, fill=blue!10, minimum size=1.2cm] (o1) at (0, 0) {$\psi_p$};
        \node[circle, draw, thick, fill=blue!10, minimum size=1.2cm] (o2) at (3, 0) {$\psi_q$};
        \node[circle, draw, thick, fill=blue!10, minimum size=1.2cm] (o3) at (6, 0) {$\psi_r$};
        \node[circle, draw, thick, fill=blue!10, minimum size=1.2cm] (o4) at (9, 0) {$\psi_s$};
        \node[rectangle, rounded corners, draw, thick, fill=red!10, minimum size=1.2cm] (a1) at (1.5, -3) {$L_1$};
        \node[rectangle, rounded corners, draw, thick, fill=red!10, minimum size=1.2cm] (a2) at (4.5, -3) {$L_2$};
        \node[rectangle, rounded corners, draw, thick, fill=red!10, minimum size=1.2cm] (a3) at (7.5, -3) {$L_3$};
        \draw[->, thick, color=gray!80] (o1) -- node[left=1mm, pos=0.4, text=black] {$B_{pq}^{L_1}$} (a1);
        \draw[->, thick, color=gray!80] (o2) -- (a1);
        \draw[->, thick, color=gray!80] (o2) -- node[right=1mm, pos=0.3, text=black] {$B_{qr}^{L_2}$} (a2);
        \draw[->, thick, color=gray!80] (o3) -- (a2);
        \draw[->, thick, color=gray!80] (o3) -- node[right=1mm, pos=0.4, text=black] {$B_{rs}^{L_3}$} (a3);
        \draw[->, thick, color=gray!80] (o4) -- (a3);
        \node at (4.5, 1.5) {\textbf{Orbital Nodes} $\mathcal{V}_O \quad \mathbf{x}_p = [h_{pp}, \|h_{p}\|_2]$};
        \node at (4.5, -4.5) {\textbf{Auxiliary Interaction Nodes} $\mathcal{V}_A \quad \mathbf{h}_L = \mathbf{0}$};
    \end{tikzpicture}
    \caption{Bipartite architecture induced by the Cholesky factorization of
    the ERI tensor. Orbital nodes (blue circles) exchange information exclusively
    through auxiliary interaction nodes (red squares) weighted by the Cholesky
    vectors $B^L_{pq}$; no direct orbital-to-orbital edges exist.}
    \label{fig:bipartite_arch}
\end{figure}

\subsection{Factorized Message Passing}

Message passing is structurally constrained by the bipartite topology. Orbital
state representations $\mathbf{x}^{(t)}$ are transmitted to the auxiliary nodes
by contracting over the pairwise Cholesky weights:
\[
  \mathbf{m}_L^{(t)} = \sum_{p,q} B^L_{pq}\,\phi\!\left(
    \mathbf{x}_p^{(t)},\, \mathbf{x}_q^{(t)}\right)
\]
where $\phi$ computes the element-wise product of orbital feature vectors
(see Appendix~\ref{app:math} for the full tensor algebra). Auxiliary nodes
process these aggregated messages via an MLP, updating their latent state
$\mathbf{h}_L^{(t)}$. The states are then broadcast back to the orbital nodes:
\[
  \mathbf{m}_p^{(t)} = \sum_{L,q} B^L_{pq}\,\psi\!\left(
    \mathbf{h}_L^{(t)},\, \mathbf{x}_q^{(t)}\right)
\]
yielding the residual update
$\mathbf{x}_p^{(t+1)} = \mathbf{x}_p^{(t)} + \text{MLP}(\mathbf{m}_p^{(t)})$.
By executing these contractions via dense einsum operations, the architecture
avoids explicit materialization of the $\mathcal{O}(N^4)$ edge adjacency matrix.

\subsection{Computational Complexity}

\begin{figure}[ht]
    \centering
    \includegraphics[width=0.65\linewidth]{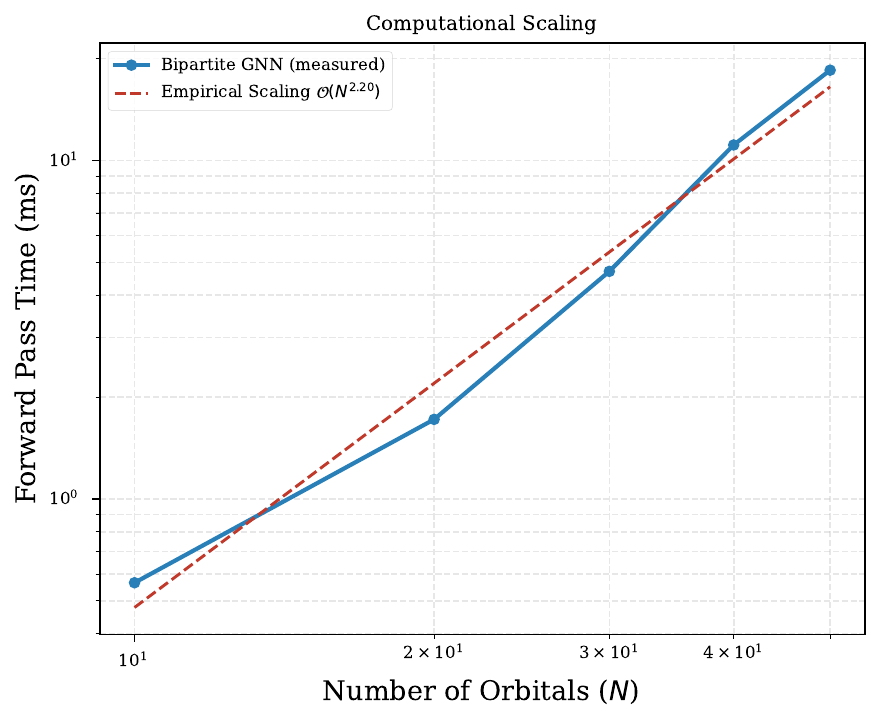}
    \caption{Empirical computational scaling of the factorized bipartite message
    passing. The forward pass achieves an empirical scaling of
    $\mathcal{O}(N^{2.20})$ on CPU, avoiding explicit $\mathcal{O}(N^4)$ edge
    materialization.}
    \label{fig:scaling}
\end{figure}

Dense tensor contractions over the Cholesky index structure yield a forward pass
that scales sub-cubically in practice. Benchmarking on CPU with orbital counts
$N \in \{10, 20, 30, 40, 50\}$ and $N_{aux} = 2N$ yields an empirical scaling
exponent of $\mathcal{O}(N^{2.20})$ (Figure~\ref{fig:scaling}). For $N = 50$
active orbitals, inference time remains below 20 ms on standard consumer
hardware, indicating computational viability for scalable implementations.

\section{Experiments}

Our primary objective is not state-of-the-art quantum chemistry accuracy on
large molecular datasets, but rather the evaluation of whether factorized
operator structure improves representation learning in orbital graph
architectures. To this end, we evaluate the bipartite Cholesky network on an
established FCI benchmark under controlled, reproducible conditions.

\subsection{Dataset}

We evaluate the architecture on the PennyLane diatomic benchmark
\citep{bergholm2022pennylane}, comprising 132 geometries across 6 diatomic
molecules (CO, HF, Li$_2$, LiH, N$_2$, O$_2$, 22 bond-length configurations
each). The STO-3G basis is used throughout. One-body integrals $h_{pq}$ and
density-fitted Cholesky vectors $B^L_{pq}$ were extracted via PySCF with a
Cholesky screening threshold of $10^{-6}$ a.u. Open-shell species (O$_2$,
triplet ground state) were treated at the ROHF level; all other molecules were
treated with RHF. The target variable is the FCI correlation energy
$\Delta E_{corr} = E_{FCI} - E_{HF}$.

\subsection{Baselines}

We benchmark against three internally implemented reference configurations,
trained on identical data splits and targets to enable controlled comparison:
(i)~an MLP acting on the flattened $h_{pq}$ matrix; (ii)~a DeepSet network
\citep{gilmer2017neural} with uncoupled orbital node embeddings that sums orbital
features without any message passing; and (iii)~a compressed orbital GNN that
follows the architecture of \citet{welborn2018transferability} and
\citet{chuiko2025predicting}, constructing edge features from Coulomb ($J$) and
exchange ($K$) matrix traces without explicit access to the Cholesky auxiliary
dimension. All baselines use the same hidden dimension, training schedule, and
Huber loss as the proposed model.

\subsection{Main Results and Ablation}

Table~\ref{tab:main_results} reports the five-fold cross-validation MAE for
in-distribution geometries, with out-of-fold predictions accumulated across all
folds to eliminate optimistic bias from single train/test splits. The bipartite
Cholesky network achieves an OOF MAE of $0.0296 \pm 0.0176$ Ha, a substantial
reduction relative to the compressed orbital GNN at $0.51 \pm 0.08$ Ha.

To validate the structural contribution of the auxiliary interaction nodes, we
performed an ablation in which the bipartite message-passing loop is replaced by
a homogeneous deep-set aggregation over orbital embeddings (using a dedicated
\texttt{OrbitalOnlyGNN} class with no shared parameters). Removing access to
the Cholesky-structured auxiliary channels increased the error to
$0.0665 \pm 0.0173$ Ha---a $2.2\times$ degradation---suggesting that the
bipartite pathway encodes pairwise correlation structure that is not recoverable
from one-body features alone.

\begin{table}[h]
  \caption{Five-fold cross-validation MAE on FCI correlation energy prediction.
  $^\dagger$GNN$_G$ targets \emph{total} FCI energy (not $\Delta E_{corr}$) on
  an 80/20 geometry split with the largest per-molecule outlier excluded;
  reported as per-molecule average across five molecules \citep{carrasquilla2025gnn}.
  Direct numerical comparison should account for these protocol differences.}
  \label{tab:main_results}
  \centering
  \begin{tabular}{lll}
    \toprule
    Model & Graph Structure & MAE (Hartree) \\
    \midrule
    MLP (Flattened $h_{pq}$)     & None                   & $1.02 \pm 0.15$ \\
    DeepSets                     & Uncoupled Orbital Nodes & $0.85 \pm 0.12$ \\
    Compressed Orbital GNN       & Compressed $J, K$ edges & $0.51 \pm 0.08$ \\
    GNN$_G$ \citep{carrasquilla2025gnn}$^\dagger$
                                 & Fock/$J$/$K$ Orbital Graph & ${\sim}0.55$ \\
    \textbf{Bipartite-Chol (Ours)} & \textbf{Factorized Bipartite} & $\mathbf{0.0296 \pm 0.0176}$ \\
    \midrule
    Ablation: No Aux.\ Nodes     & Orbital Homogeneous    & $0.0665 \pm 0.0173$ \\
    \bottomrule
  \end{tabular}
\end{table}

\begin{figure}[ht]
    \centering
    \includegraphics[width=\linewidth]{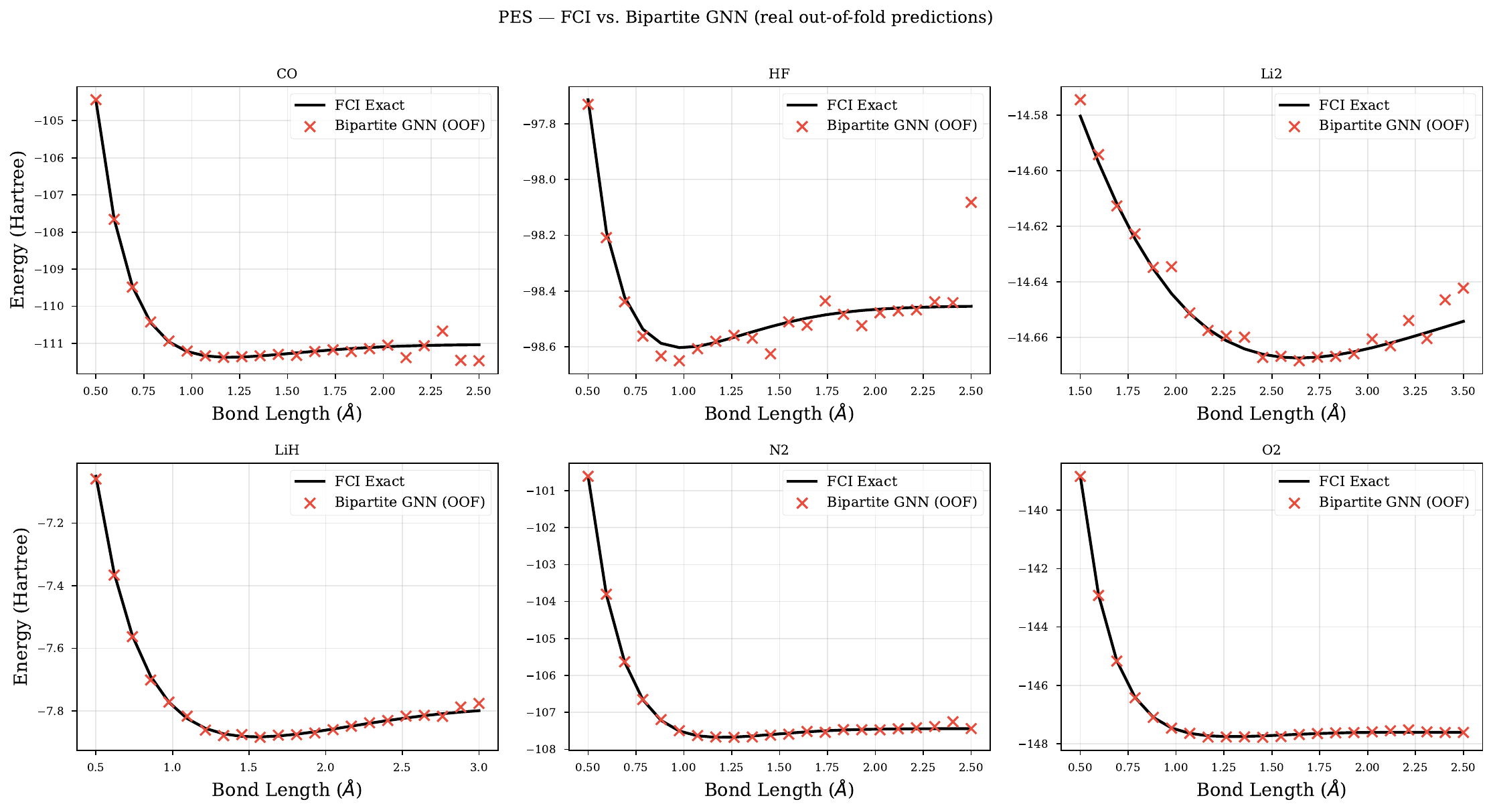}
    \caption{Potential energy surfaces for the six diatomic molecules. Solid
    lines represent FCI reference energies; crosses are genuine out-of-fold
    predictions of the bipartite GNN accumulated across all five CV folds. No
    prediction was produced by a model that had seen the corresponding geometry
    during training.}
    \label{fig:pes_curves}
\end{figure}

Figure~\ref{fig:pes_curves} shows the predicted potential energy surfaces (PES)
for all six molecules. Predictions represent genuine out-of-fold estimates from
the five-fold CV procedure; no prediction in the figure was produced by a model
trained on that geometry. The model tracks the equilibrium region faithfully
across all species and exhibits larger residuals near compressed geometries
(bond length $< 0.75$ \AA), where strong-correlation effects become significant
and the single-reference Hartree-Fock reference is least adequate.

\subsection{Leave-One-Molecule-Out Transfer Analysis}
\label{sec:lomo}

To probe the representational boundaries of the model, we performed a
leave-one-molecule-out (LOMO) validation, training on five species and
evaluating zero-shot on the held-out molecule. Early stopping was performed on
a validation split drawn from the \emph{training} molecules; the held-out
species was evaluated only once using the best checkpoint, ensuring no
information from the test molecule influenced model selection.

\begin{figure}[ht]
    \centering
    \includegraphics[width=0.65\linewidth]{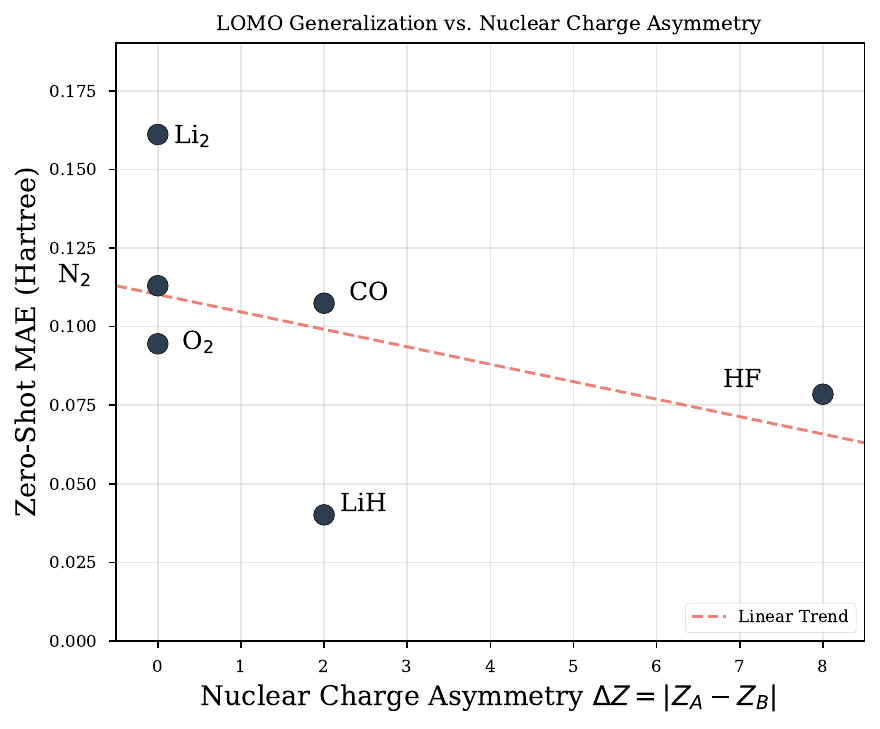}
    \caption{Zero-shot generalization error (LOMO-CV) as a function of nuclear
    charge asymmetry $\Delta Z = |Z_A - Z_B|$. The absence of a consistent
    monotonic trend indicates that $\Delta Z$ alone does not explain
    transferability. LiH (low $\Delta Z$, low error) and Li$_2$ (zero $\Delta Z$,
    highest error) illustrate that the dominant factor is the orbital-structural
    dissimilarity of the held-out species from the training distribution.}
    \label{fig:lomo}
\end{figure}

Zero-shot MAE ranges from $0.040$ Ha (LiH) to $0.161$ Ha (Li$_2$),
a factor of four variation across species. The results are shown in
Figure~\ref{fig:lomo} and Table~\ref{tab:lomo}.

\begin{table}[h]
  \caption{LOMO zero-shot MAE by held-out molecule.}
  \label{tab:lomo}
  \centering
  \begin{tabular}{lcc}
    \toprule
    Molecule & $\Delta Z = |Z_A - Z_B|$ & Zero-Shot MAE (Ha) \\
    \midrule
    LiH  & 2 & 0.0401 \\
    HF   & 8 & 0.0785 \\
    O$_2$  & 0 & 0.0945 \\
    CO   & 2 & 0.1074 \\
    N$_2$  & 0 & 0.1130 \\
    Li$_2$ & 0 & 0.1611 \\
    \bottomrule
  \end{tabular}
\end{table}

Contrary to a symmetry-based prediction in which homonuclear species (sharing
inversion symmetry with the training molecules) would transfer most readily, the
three homonuclear molecules span both extremes of the generalization spectrum:
O$_2$ is mid-range ($0.094$ Ha) while N$_2$ ($0.113$ Ha) and Li$_2$ ($0.161$ Ha)
exhibit the largest transfer errors. The simple $\Delta Z$ correlation does not
hold: Figure~\ref{fig:lomo} shows no consistent trend with nuclear charge
asymmetry.

The data are better described by the orbital-structural similarity of the
held-out species to the training distribution. LiH transfers most readily
because both of its constituent atomic environments appear independently in the
training set: lithium chemistry is represented by Li$_2$ and hydrogen chemistry
by HF, giving the auxiliary nodes a prior over both atomic interaction vertices.
Li$_2$, by contrast, is the hardest case despite being homonuclear. Its bonding
is dominated by the overlap of two diffuse 2s lithium orbitals, forming a
weakly-bound $\sigma$ system with no close structural analogue among the five
training molecules (CO, HF, LiH, N$_2$, O$_2$), all of which involve tighter
2p orbital bonding or mixed-character $\sigma$-$\pi$ systems. The auxiliary
node representations, trained entirely on these harder-bond chemistries,
appear to encode an orbital-interaction prior that does not transfer well to
the diffuse Li$_2$ case.

\subsection{Permutation Invariance}

We verify that arbitrary permutation of the orbital indexing $(p, q)$ within
the $h_{pq}$ and $B^L_{pq}$ input tensors yields invariant energy predictions
($|\Delta E| < 10^{-5}$ Ha). The test uses a physically valid symmetric
Cholesky tensor ($B^L_{pq} = B^L_{qp}$) consistent with the ERI supermatrix
structure, confirming that the architecture preserves orbital-label parity.

\section{Discussion}

The performance advantage of the bipartite architecture stems from its structural
consistency with the operator factorization. Standard node-to-node GNNs struggle
to represent four-index tensors without combinatorial scaling; by defining the
auxiliary Cholesky basis functions as explicit graph nodes, the architecture
processes pairwise orbital correlation in a manner consistent with the structure
of the Coulomb metric, without ever materializing the $\mathcal{O}(N^4)$ ERI.

The LOMO analysis reveals that transferability is governed by the
orbital-structural similarity of the held-out molecule to the training
distribution. Species whose constituent atomic interaction environments are
represented in the training set---such as LiH, whose atomic building blocks
appear separately in Li$_2$ and HF---transfer most readily. Species with
qualitatively distinct bonding characters---such as Li$_2$, whose diffuse
$\sigma$-bonded 2s system has no close analogue in the training set---exhibit
the largest zero-shot errors. This suggests that the auxiliary node
representations learn an orbital-interaction prior that is anchored to the
chemical diversity of the training molecules rather than to any single symmetry
descriptor.

The $2.2\times$ degradation observed in the ablation study confirms that the
bipartite message-passing pathway is responsible for the bulk of the predictive
improvement, rather than the depth of the energy readout head or the choice of
one-body node features alone.

\section{Limitations}

Our empirical validation is constrained to six diatomic molecules in a minimal
STO-3G basis, which limits the diversity of orbital structures encountered during
training. The current architecture relies on a single-reference Hartree-Fock
reference (RHF or ROHF), precluding direct evaluation on strongly
multi-reference systems or excited states. The DeepSets and Compressed Orbital GNN baselines in Table~\ref{tab:main_results} are internally implemented reference configurations. The closest externally published result is the concurrent GNN$_G$ of
\citet{carrasquilla2025gnn}, which targets total FCI energy rather than
correlation energy and reports a per-molecule MAE of ${\sim}0.55$ Ha on an
80/20 split over the same benchmark; our 5-fold OOF MAE of $0.0296$ Ha on the
$\Delta$-ML correlation target represents a ${\sim}15{\times}$ reduction in
absolute error under a stricter evaluation protocol.

\section{Conclusion}

We present a bipartite graph neural network motivated by the tensor factorization
of the molecular Hamiltonian. By preserving the Cholesky structure of the ERI
tensor as explicit auxiliary graph nodes rather than compressing it into scalar
edge features, the architecture achieves a substantial reduction in correlation
energy prediction error while maintaining sub-cubic practical scaling. LOMO
validation suggests that zero-shot generalization tracks orbital-structural
similarity to the training distribution. Factorized operator representations may
provide a general design principle for structuring geometric deep learning
architectures in quantum chemistry, particularly as larger and more chemically
diverse orbital datasets become available.

\paragraph{Code Availability}
All experiments, model code, and data generation scripts are available at
\url{https://github.com/maestroK/bipartite-cholesky-gnn}.

\bibliographystyle{plainnat}
\bibliography{references}

\newpage
\appendix

\section{Extended Mathematical Formulation}
\label{app:math}

This appendix details the theoretical mappings connecting the quantum many-body
Hamiltonian to the factorized bipartite graph architecture, as well as the formal
tensor operations comprising the message-passing framework.

\subsection{The Coulomb Metric and Positive Semi-Definiteness}

The non-relativistic electronic Hamiltonian in second quantization is defined as:
\begin{equation}
    \hat{H} = \sum_{pq}^N h_{pq} \hat{a}_p^\dagger \hat{a}_q +
    \frac{1}{2}\sum_{pqrs}^N g_{pqrs} \hat{a}_p^\dagger \hat{a}_q^\dagger
    \hat{a}_r \hat{a}_s
\end{equation}
where $h_{pq}$ represents the one-body core Hamiltonian and $g_{pqrs}$ the
two-electron repulsion integrals (ERI). The ERI tensor evaluates the Coulomb
interaction between two orbital pair densities:
\begin{equation}
    g_{pqrs} = \int \!\!\int \phi_p^*(\mathbf{r}_1) \phi_q(\mathbf{r}_1)
    \frac{1}{|\mathbf{r}_1 - \mathbf{r}_2|}
    \phi_r^*(\mathbf{r}_2) \phi_s(\mathbf{r}_2)\, d\mathbf{r}_1\, d\mathbf{r}_2
\end{equation}
By defining a composite index mapping $I = (pq)$ and $J = (rs)$, the four-index
ERI tensor can be flattened into a symmetric supermatrix $G_{IJ} \equiv g_{pqrs}$.
The ERI supermatrix induced by the Coulomb kernel is positive semi-definite (PSD).

\subsection{Cholesky Factorization of the Interaction Vertex}

The PSD nature of $\mathbf{G}$ guarantees the existence of a Cholesky
decomposition. We compute the incomplete Cholesky factorization
$\mathbf{G} \approx \mathbf{L}\mathbf{L}^T$, subject to a truncation
tolerance $\epsilon$:
\begin{equation}
    G_{IJ} \approx \sum_{L=1}^{N_{aux}} L_{I,L}\, L_{J,L}
\end{equation}
where $\mathbf{L}$ is lower-triangular of dimension $N^2 \times N_{aux}$.
Restoring the composite index $I = (pq)$ defines the three-index Cholesky
tensor $B^L_{pq} \equiv L_{(pq),L}$, recovering:
\begin{equation}
    g_{pqrs} \approx \sum_{L=1}^{N_{aux}} B^L_{pq}\, B^L_{rs}
\end{equation}
This factorization reduces the structural dimension of the interaction space from
$\mathcal{O}(N^4)$ to $\mathcal{O}(N^2 N_{aux})$. Because linear dependencies in
the auxiliary basis scale approximately as $N_{aux} \sim N$ in practice
\citep{koch2003reduced}, the theoretical scaling is broadly bounded by
$\mathcal{O}(N^3)$.

\subsection{Tensor Algebra of Bipartite Message Passing}

Let $\mathbf{x}_p^{(t)} \in \mathbb{R}^H$ be the latent representation of
orbital node $p$ at layer $t$, and $\mathbf{h}_L^{(t)} \in \mathbb{R}^H$ the
representation of auxiliary node $L$. Initial features are
$\mathbf{x}_p^{(0)} = [h_{pp}, \|h_p\|_2]\mathbf{W}_{emb}$ and
$\mathbf{h}_L^{(0)} = \mathbf{0}$.

The Orbital-to-Auxiliary message aggregates pairs of orbitals through
interaction channel $L$:
\begin{equation}
    \mathbf{m}_L^{(t)} = \sum_{p=1}^N \sum_{q=1}^N B^L_{pq}
    \left( \mathbf{x}_p^{(t)} \odot \mathbf{x}_q^{(t)} \right)
    \label{eq:o2a}
\end{equation}
where $\odot$ denotes the element-wise Hadamard product in feature dimension $H$.
Note that this is equivalent to the outer-product formulation
$\phi(\mathbf{x}_p, \mathbf{x}_q)$ in the main text; the summation contracts
the orbital indices while preserving the feature dimension. Auxiliary node states
are updated via:
\begin{equation}
    \mathbf{h}_L^{(t+1)} = \mathbf{h}_L^{(t)} +
    \sigma\!\left( \mathbf{W}_A^{(t)} \mathbf{m}_L^{(t)} \right)
    \label{eq:aux_update}
\end{equation}

The Auxiliary-to-Orbital message broadcasts auxiliary states back:
\begin{equation}
    \mathbf{m}_p^{(t)} = \sum_{L=1}^{N_{aux}} \sum_{q=1}^N B^L_{pq}
    \left( \mathbf{h}_L^{(t+1)} \odot \mathbf{x}_q^{(t)} \right)
    \label{eq:a2o}
\end{equation}

Orbital states are residually updated:
\begin{equation}
    \mathbf{x}_p^{(t+1)} = \mathbf{x}_p^{(t)} +
    \sigma\!\left( \mathbf{W}_O^{(t)} \mathbf{m}_p^{(t)} \right)
    \label{eq:orb_update}
\end{equation}

By executing Equations~\eqref{eq:o2a}--\eqref{eq:orb_update} via dense einsum
tensor contractions, the architecture circumvents explicit allocation of an
$\mathcal{O}(N^4)$ edge adjacency matrix.

\section{Justification of the $\Delta$-Machine Learning Formulation}
\label{app:delta_ml}

Predicting the absolute total energy $E_{FCI}$ using a single scalar output
exposes the network to a severe energy-scale imbalance. The total energy
decomposes as $E_{FCI} = E_{HF} + \Delta E_{corr}$. In our dataset, the
variance of $E_{FCI}$ across six disparate molecular species spans
$\mathcal{O}(10^2)$ Hartree (from $-7.8$ Ha for LiH to $-148$ Ha for O$_2$),
while $\Delta E_{corr}$ spans only $\mathcal{O}(10^{-1})$ Hartree. Because
the input graph uses strictly electronic integrals ($h_{pq}$, $B^L_{pq}$),
predicting $E_{FCI}$ requires the network to implicitly reconstruct the nuclear
repulsion $V_{nn}$ from orbital features alone, dominating the gradient and
masking the many-body objective. By adopting $\Delta$-ML
\citep{ramakrishnan2015big} and defining the loss on $\Delta E_{corr}$, we
isolate the objective to the many-body scattering processes parameterized by
the auxiliary interaction nodes.
\end{document}